\documentclass[twocolumn,showpacs,preprintnumbers,amsmath,amssymb,pre]{revtex4}
\usepackage{graphicx}% Include figure files
\usepackage{dcolumn}% Align table columns on decimal point
\usepackage{bm}% bold math

\newcommand{\be}{\begin{eqnarray}}
\newcommand{\ee}{\end{eqnarray}}
\newcommand{\tri}{\mbox{\scriptsize $\triangle$}}
\begin{document}

\preprint{cond-mat/0305515}

\title{Recurrent Biological Neural Networks:\\ The Weak and 
Noisy Limit}% Force line breaks with \\

\author{Patrick D. Roberts}
 \email{robertpa@ohsu.edu}
\homepage{http://www.ohsu.edu/nsi/faculty/robertpa}
\affiliation{%
Neurological Sciences Institute \\
Oregon Health \& Science University\\
505 N.W. 185th Avenue\\
Beaverton, OR 97006
}%

\date{\today}% It is always \today, today,

\begin{abstract}
A perturbative method is developed for calculating the effects of recurrent synaptic interactions between neurons embedded in a network. A series expansion is  constructed that converges for networks with noisy membrane potential  and weak synaptic connectivity. The terms of the series can be interpreted as loops of interactions between neurons, so the technique is called a loop-expansion. A diagrammatic method is introduced that allows for construction of analytic expressions  for the parameter dependencies of the spike probability function and correlation functions.   An analytic expression is obtained to predict the effect of the surrounding network on a neuron during an intracellular current injection. The analytic results are compared with simulations to test the range of their validity and significant effects of the the recurrent connections in network are accurately predicted by the loop-expansion. 
\end{abstract}

\pacs{87.18.Sn,87.19.La,75.10.Nr}% PACS, the Physics and Astronomy
\maketitle

\section{\label{sec:intro}Introduction}
Recurrent connectivity is a ubiquitous feature of biological neural
networks, particularly in the central nervous system of vertebrates. 
This recurrent structure has made mathematical analysis difficult
because of the nonlinearities that arise when the action of a
particular neuron affects itself through its action on neighboring
neurons \cite{McCulloch43}.  In addition, interpretation of electrophysiological data can be
misleading because the activity of neighboring neurons to the
recording site are typically affected by the experimental stimulus \cite{Amit97,Tsodyks95,Mattia02,BenYishai95}. 

Mathematical methods that can quantify the effects of neighboring
neurons on the response of a single neuron embedded in a network would
greatly improve the predictions of neural network models and ease the
comparison with system-level electrophysiological recordings. The development of theoretical methodology has led to the study of idealized systems such as syn-fire chains \cite{Abeles91}, networks with correlated inputs \cite{Svirskis00,Salinas02}, and large networks of identical neurons \cite{Nykamp00,Haskell01}. The goal of this project is
to develop analytic methods to improve quantitative tests of electrophysiological hypotheses
in order to set stricter limits on our theories of brain function.

Quantitative tests require mathematical methods that separate
different dynamic modes and can be refined to dynamics of the
particular system of interest.  We shall separate the dynamical modes
of recurrence into \emph{coherent} and \emph{asynchronous} modes \cite{Abbott93,vanVreeswijk00}.  The
coherent mode is characterized by strong synaptic connections that
cause the activity of one neuron to dramatically affect the activity
of another.  This dynamical mode is essential to central pattern
generators. A simple example is the half-center oscillator, consisting of a pair of symmetrically coupled neurons that are
strongly coupled by inhibitory synaptic currents.  When one neuron
fires, the other neuron is silent.  If each neuron expresses intrinsic
conductances to terminate a burst, and to induce firing once they are
released from inhibition, then the neuronal pair will oscillate.  The
coherent mode is quantified by the correlation function that is
nearly $\pm 1$, for  synaptic couplings that are either excitatory (+1)
or inhibitory (-1).

The present article investigates the effects of recurrence in the
asynchronous state when the joint-correlation between neurons is small but 
does not vanish.  We
quantify the parametric dependence of neural-response variables in
the asynchronous state on the synaptic strength, the spike probability,
and the presence of noise in the system.  We also show that the
influence of neurons on their neighbor can be considerable in the
asynchronous state, suggesting that such corrections should be included
in our predictions about the results of physiological experiments,
even when no coherent state dynamics are obvious.

Another interesting result found in this analysis is that, 
under physiological conditions found in the vertebrate central 
nervous system, the effects of recurrent connectivity propagates across 
few synaptic contacts. Thus, if distance is measured by the number of synaptic junctions separating neurons, effects of distant neurons can be 
safely neglected in models of biological neural networks. Thus, large-scale simulations of biological neural networks could be simplified to smaller networks that produce quantitatively similar results.

In the following two subsections, we introduce our neural modeling
methods and establish our notation.  The next section investigates the
effects of recurrence on the simple model of a pair of neurons.  We
show how to calculate the correction to the spike
probability and membrane potential due to recurrent synaptic connections.  The method considers recurrent
connections as a perturbation to the background activity of the mean field result \cite{Treves93}.  The terms of the perturbation series are weighted by powers of the synaptic strength multiplied
by the inverse of the noise in the system.  Thus, the perturbation expansion
converges quickly for systems that are noisy and weakly coupled. 

We compare the results of the perturbation expansion to numerical
simulations to test the validity of the method and quantify the limits
of this approach.  Other experimentally measurable quantities are also
calculated: the response to a stimulus, the auto-correlation function,
and the joint-correlation function.  Section \ref{sec:multiple}
generalizes our perturbation method to systems with many neurons and
shows that in the asynchronous mode, only close neighbors contribute
significantly.  The final section discusses the applications of these
analytic methods to biological systems.

\subsection{\label{sec:SRN}Network of spike response neurons}

The model neurons used in this study are based on the Spike-Response
Model \cite{Gerstner93}.  The basic idea is to construct a kernel
function that describes the response of a  biological neuron to
spikes.  The membrane time constant and other intrinsic properties of
neurons are contained in the response function \cite{Gerstner95}.  In the following, we
restrict the model to respond linearly to presynaptic spikes with a
simple exponential decay represented by the \emph{postsynaptic-potential (PSP) kernel}, $\epsilon(t)$.  If $t_{pre}$ is the time of the
presynaptic spike, then $\epsilon(t) = a \exp [-a(t-t_{pre})] \mbox{ for } t
\geq t_{pre}$, otherwise $\epsilon(t) = 0$. The parameter $a$ is used
to describe the membrane time constant, and the kernel is normalized
such that $\int\epsilon(t) \, dt=1$. This choice of kernel implies that the model is formally equivalent  \cite{Gerstner98a} to integrate-and-fire models\cite{Jack75,Stein67b}. These kernels could also contain a synaptic delay, but in the examples we will assume that the delay is negligible.

\begin{figure}
\includegraphics[width=3.4in]{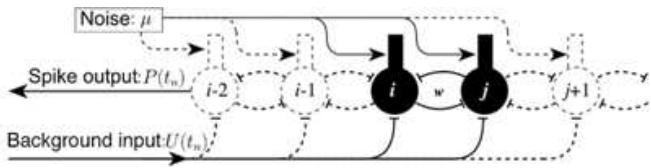}% Here is how to import EPS art
\caption{\label{fig:model} Recurrent network model.  Spike response
neurons are coupled by recurrent synaptic connections with weight $w$. The spike probability, $P(t_n)$, of each model neuron is a function of the synaptic connections, a background input, $U(t_n)$, and internal noise, $\mu$. }
\end{figure}

Spikes are generated in each model neuron $i$ by a probability
function, $P_{i}(t)$, that is dependent on a variable called the {\it
membrane potential}, $V_{i}(t)$.  The membrane potential is the
sum of all synaptic inputs.  It is convenient to construct our
probability function in discrete time where, $t_{n} = n \tri t$, $n\in \mathbb{Z}$ and
$\tri t$ is on the order of one millisecond.  The probability of a spike
generated in neuron $i$ during time step $t_{n}$ is given by a
threshold function of the membrane potential,
\be
P_{i}(t_{n}) = 
P_{i}(V_{i}(t_{n})) = \frac{1}{1+\exp[-\mu_i(V_{i}(t_{n})-\theta_i)]}
\label{eq:probfnct}
\ee
where $\theta_i$ is the spike threshold and $\mu_i$ parametrizes the noise ($\mu_i \sim$ 1/noise) of neuron $i$. The noise arises from ion channel fluctuations that are internal to each neuron, and from synaptic noise that arises from synaptic connections not explicitly included in the model.

A presynaptic spike evoked by another neuron, $j$, will contribute to the 
membrane potential $V_{i}(t_{n})$ with a PSP 
kernel
weighted by the synaptic weight, $w_{ij}$. In addition to synaptic 
input, the membrane potential of neuron $i$ is also given a bias (or background input), 
$U_{i}(t_{n})$, that provides a spontaneous discharge rate or driving input 
to the neuron $i$. If $S_{j}^{pre}(t_{m})$ represents a spike train ($S_{j}^{pre}(t_{m})$ = 1or 0 at each time step) from neuron $j$, then the membrane potential of neuron $i$ is,
\be
V_{i} (t_{n}) = U_{i} (t_{n}) + 
              \sum_{m, j} w_{ij}\epsilon(t_{n}-t_{m})S_{j}^{pre}(t_{m}).
\ee
The $j$-sum is over all neurons that have synaptic contact onto 
neuron $i$. To compute the ensemble average of the membrane potential, 
we use the probability functions of the presynaptic neurons,
\be
\langle V_{i} (t_{n})\rangle  = U_{i} (t_{n}) + \sum_{m, j} w_{ij}\epsilon(t_{n}-t_{m})P_{j}(t_{m})
\label{eq:mempot}
\ee
because $P_{j}(t_{m}) = \langle S_{j}^{pre}(t_{m})\rangle$.
Since the spike probability of neuron $j$ is functionally dependent 
on the spike probability of neuron $i$, via its membrane potential, 
then in recurrent circuits $V_{i} (t_{n})$ will appear nonlinearly on both sides 
of the equation and we must seek alternative methods to solve for it.

\subsection{\label{sec:fixedpt}Fixed-point behavior and time scales}
The neurons of neural systems have been found to have preferred spike rates
due to adaptive mechanisms that drive these systems to these 
rates and help maintain stability of the overall system \cite{Turrigiano99,Bell97b} . Both 
theoretical \cite{Song00,vanRossum00,Roberts00a} and experimental studies \cite{Turrigiano99}  have suggested that the 
establishment of a fixed point in the neural dynamics is on a time scale 
that is much longer than the dynamics driven by spike responses. The 
presence of a fixed point allows one to choose a mean membrane 
potential to expand the probability function in Eq.\ 
(\ref{eq:probfnct}). Let $\hat{p}$ be the stable fixed point of the 
long time scale dynamics:
\be
P_{i}(t_{n}) \rightarrow \hat{p}_{i} = P_{i}(\hat{V}_{i})
\mbox{  as  } t_{n} \rightarrow \infty
\ee
By expanding the probability function about this fixed point, we 
can extract a linear relation to expose the membrane-potential 
function on the right side of Eq.\ (\ref{eq:mempot})
\be
P_{j}(t_{n}) = \hat{p}_{j} + \mu_j (\hat{p}_{j} - 
\hat{p}^{2}_{j})(V_{j}(t_{n})-\hat{V_{j}})+\cdots
\label{eq:expanj}
\ee
where $\hat{V_{j}}$ is the membrane potential at the fixed point.
In the following, we will assume that the system is near the fixed point, $\hat{p}$, that is independent of time. This assumption will simplify the analysis and, in one example, we will allow the background input, $U_{i} (t_{n})$, to vary in time. 

In the next section we will pursue this expansion by simplifying our
network to the case with only two synaptically coupled neurons to illustrate the method. However, the technique generalizes to networks of multiple neurons as will be shown Section III.

\section{\label{sec:pair}Pair of coupled neurons}
Let $w_{ij} \in$ $\mathbb{R}$ be the weight (efficacy) of a synaptic 
connection from neuron $j$ onto neuron $i$. For our 2-neuron network, 
let $U_{i}(t_{n})$ be the background membrane potential of neuron $i$, 
the membrane 
potential in the absence of synaptic connections with neuron $j$. Here, the symbols $i$ and $j$ are fixed labels for each of the two neurons so that we may easily generalize the method to larger networks in the following sections. The 
ensemble average membrane potential of neuron $i$ is given by 
Eq. (\ref{eq:mempot}). However, since $P_{j}(t_{m})$ is dependent on 
the activity of neuron $i$ due to the recurrent synaptic connection, 
we must expand about the background membrane potential of neuron $j$ as in 
Eq. (\ref{eq:expanj}). The membrane potential of neuron $j$ is also 
dependent on the spike activity of neuron $i$, so that we now have,
\begin{widetext}
\be
\langle V_{i} (t_{n})\rangle  = U_{i} (t_{n})  
+ w_{ij}\hat{p}_{j}\epsilon*[1+\mu (1-\hat{p}_{j})( U_{j}(t_{m}) - \hat{V_{j}}) ]
+\mu w_{ij}\hat{p}_{j}(1-\hat{p}_{j})w_{ji}
\epsilon^{(2)} * P_{i}(t_{n})
\label{eq:mempot2}
\ee
\end{widetext}
where we have introduced the notation for the multiple convolutions,
$\epsilon^{(K)}*f(t_n)$ =
$\sum_{i_{1}}\cdots\sum_{i_{K}}\epsilon(t_{i_{K-1}}-t_{i_{K}})\cdots
\epsilon(t_{n}-t_{i_{1}})f(t_{i_1})$. The last term in Eq.\ \ref{eq:mempot2} represents the recurrent effects of neurons $i$'s activity on itself via neuron $j$. Expanding the spike probability
function of neuron $i$, $P_{i}(t_{n})$, generates another term with a
factor that contains the spike probability function of neuron $j$.

For simplicity, let $\hat{p}_{i}=\hat{p}_{j}=\hat{p}$, $\mu_i=\mu_j=\mu$,  $\theta_i=\theta_j=\theta$, and collect the terms in powers of the synaptic weight, $w_{ij}=w_{ji}=w$, 
to arrive at an expression for the average membrane potential,
\begin{widetext}
\be
\langle V_{i} (t_{n})\rangle  = U_{i} (t_{n})  
+ w\hat{p}\sum_{K=1}^{\infty} 
(w\mu \hat{p}(1- \hat{p}))^{K-1}
[1 + \mu(1-\hat{p})
\epsilon^{(K)} * (U_{a}(t_{n}) - \hat{V_{a}})]
\label{eq:mempotexp}
\ee
\end{widetext}
where
\be
a = \left\{\begin{array}{ll}
i & \mbox{if $K$ is even}\\
j & \mbox{if $K$ is odd}
\end{array}
\right\}
\ee
We now have an expression that is dependent on the background activity of 
either neuron, that is, we may calculate the effect of the recurrent 
connections on any specific neuron to an accuracy of order 
$w\mu\hat{p}(1- \hat{p})$. 

The spike probability can also be expressed as a power series by
substituting the expansion of the average membrane potential (Eq.\
\ref{eq:mempotexp}) back into
the expression for the spike probability function Eq.\
(\ref{eq:expanj}):
\be
P_{i}(t_{n}) = \sum_{K=0}^{\infty} (w\mu \hat{p}(1- \hat{p}))^{K}
\epsilon^{(K)}*P^U_{a}(t_{n})
\label{eq:probloop}
\ee
where $P^U_{a}(t_{n})=P_{a}(U_a(t_{n}))$ is the background spike probability of neuron
$a$, and we have defined $\epsilon^{(0)}(t_{n}) = \delta(t_{n})$. Notice that the fixed-point, $\hat{p}$,
is still present in this expression.  If the expansion were taken
about the background spike probability, the expression is the same unless
the background spike probability is not constant.

For a sensible prediction of the spike probability function, the expansion must be bounded on the interval $[0,1]$.
Because the expansion parameter, $w\mu\hat{p}(1- \hat{p})$, is
dependent on the noise and synaptic weight, we must investigate the
conditions for convergence of this series.  The product, $\hat{p}(1-
\hat{p})\leq 0.25$ for all $\hat{p}$, so the important parameters for convergence are
the noise parameter and synaptic weight.  Since $\mu$ is small for noisy
networks, and $w$ is small for weakly coupled networks, then the perturbation series converges in the weak and noisy limit of these
network parameters provided that $\epsilon^{(K)}*P^U_{a}(t_{n})$ is bounded as $K\rightarrow\infty$.

\begin{figure}
\includegraphics[width=3.2in]{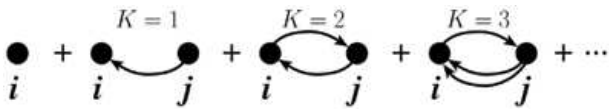}% Here is how to import EPS art
\caption{\label{fig:loops}Loop Expansion.  Each term in the series
expansion for two coupled neurons (Eq.\ \ref{eq:probloop}) 
is represented by a diagram of recurrent loops with $K$ synaptic links. }
 \end{figure}

\subsection{\label{sec:diagram}Diagrammatic methods}
The expressions for the spike probability function and average
membrane potential can become unwieldy when many synaptic connections
are involved with different weights and different neuronal types. 
Therefore, we introduce a book-keeping device based on Feynman-like
diagrams that will help keep track of the terms of the expansion \cite{Korutcheva00,Chauvet02}.
  
A {\it chain} of {\it synaptic links} is the number of synaptic connections that a 
signal passes from the starting neuron to the final neuron. For 
instance, in our example of the pair of coupled neurons, the second 
term in Fig.\ \ref{fig:loops} is a chain of 1 synaptic link, the third term is 
a chain of 2 synaptic links. Each unique chain of synaptic links contributes a term to the perturbation expansion.

For a chain of $K$ synaptic links: 
\\
$1\rightarrow 2\rightarrow\cdots i\rightarrow j\rightarrow\cdots K\rightarrow K+1$,
\\
the influence of the first neuron in the chain on the spike probability of neuron $(K+1)$ is given by a term 
containing the following factors:
\begin{enumerate}
    \item  $w_{ji}\mu_{j}\hat{p}_{j}(1-\hat{p}_{j})$ for each \emph{synaptic link}, 
    $i\rightarrow j$,

    \item  $\epsilon^{(K)}*P^U_{1}(t_{n})$,
\end{enumerate}
where 1 is the label of the first neuron in the synaptic chain.

The perturbation expansion for the average membrane potential is handled similarly. For a chain 
of $K$ synaptic links: 
\\
$1\rightarrow 2\rightarrow\cdots i\rightarrow j\rightarrow\cdots K\rightarrow K+1$,
\\
the influence of the first neuron in the chain on the membrane potential of neuron $(K+1)$ if given by a term 
containing the following factors:
\begin{enumerate}
    \item  $w_{ji}\hat{p}_{i}$ for each \emph{synaptic link}, $i\rightarrow j$,

    \item  $\mu_{i}(1-\hat{p}_{i})$ for each \emph{traverse}, $\rightarrow i\rightarrow$,

    \item  $\epsilon^{(K)}*[1 + \mu(1-\hat{p}_{1})(U_{1}(t_{n})-\hat{V_{1}})]$.
\end{enumerate}

\begin{figure*}
\includegraphics[width=6in] {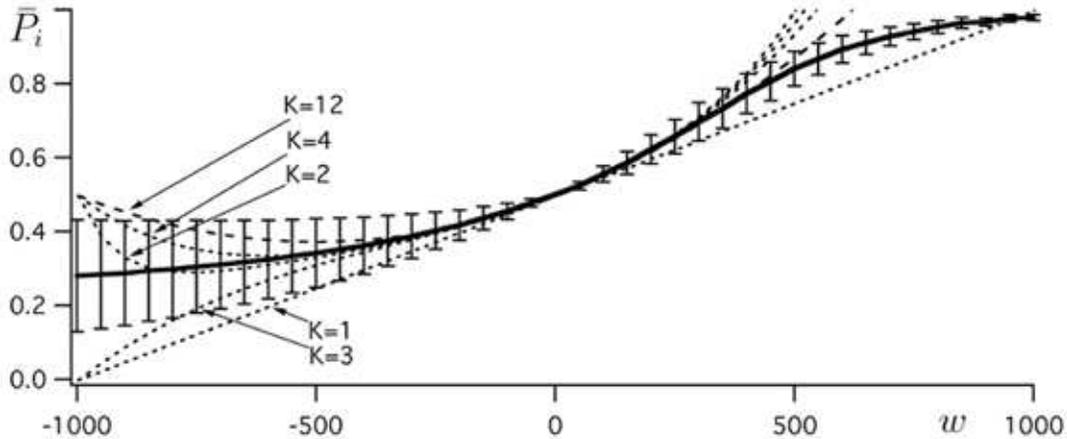}
\caption{\label{fig:probcomps} Comparison of the loop-expansion prediction with simulation for the synaptic weight dependence of the spike probability. The time average of the spike probability function (solid line)  saturates for strong synaptic weights (error bars represent 1 standard deviation). The loop-expansion prediction (dashed lines) for  $K$-terms of the expansion (Eq.\ \ref{eq:probloop}) matches the simulation result in the presence of strong effects on the spike probability from recurrent connections. Background input is set to the threshold, $U_{i}(t_{n})=U_{j}(t_{n})=\theta$, the noise parameter is $\mu = 0.002$, and the number of time steps for each simulation is $N=2\times 10^6$.}
\end{figure*}

The resultant perturbation series, called a loop expansion, approximates the effects of neighboring neurons 
by expanding in the order of loops. Fig.\  (\ref{fig:loops}) shows the graphical representation of the first four terms of Eq.\  (\ref{eq:probloop}). The first 
term is the background spike probability, and the second term represents the 
effect of the second neuron on the first if there were no recurrent 
connections. The third term is then the effect of the first neuron on 
itself via its effect on the second neuron. 
This technique generalizes to any number of neurons with 
unique synaptic weights, intrinsic noise and background spike 
probability. Analytic methods using another type of converging series have been developed previously to study population dynamics of large networks \cite{Knight00}.

\subsection{\label{sec:num}Numerical comparison}
We wish to check the range of validity of our loop expansion of our model network by comparing the parameter dependency of the loop expansions with a numerical simulation of a pair of
coupled neurons \footnote{Java simulation code used in this study can be obtained at http://www.ohsu.edu/nsi/faculty/robertpa/lab/}.   
In our simulations, we set $U_{i}(t_{n})=U_{j}(t_{n})=\theta$ so that
the background spike probability, $P^U_{i}(t_{n}) = P^U_{j}(t_{n}) =
0.5$.  With no other inputs or adaptive mechanisms in the model neurons, our
loop-expansion fixed-point will also be, $\hat{p}_{i} = \hat{p}_{j} = 0.5$. We choose this value for our comparisons because each term of the loop expansion is weighted by factor that is a power of $\hat{p}_{i}(1-\hat{p}_{i})$; a factor that is maximum for  $\hat{p}_{i}  = 0.5$. Thus, other values for $P^U_{i}(t_{n})$ and  $\hat{p}_{i}$ would allow the series to converge faster and improve the prediction of the loop expansion.

No time dependent inputs were present in the simulation, thus we compared the time average and variance of the spike probability over the sampling time to the prediction of the loop expansion. In the model neurons, recurrent spikes were generated by selecting pseudo-random numbers from a uniform distribution with a spike probability  calculated by the spike response model (Eq.\  \ref{eq:probfnct}) to simulate spike generation by a Poisson point process. This spike generation method allows us to compare the loop expansion results to simulations up to the order of variance of a uniform distribution. The sample time was $N=2\times 10^6$ time steps. The spike probability calculated at each time step was averaged over the sample time:
\be
\bar{P_i} = \frac{1}{N}\sum_{n=1}^N P_i(t_{n}),
\ee
with the variance calculated by
\be
\mbox{var}(P_i) = \sqrt{\frac{1}{N}\sum_{n=1}^N (\bar{P_i} -P_i(t_{n}))^2}.
\ee
In the figures, the variance is shown with error bars. 

Fig.\ \ref{fig:probcomps} shows the effect of synaptic strength on the theoretical
spike probability, where we set the noise parameter to $\mu = 0.002$. The
prediction for the first 12 terms of the loop-expansion is
indistinguishable from the simulation for $w \in[-900, 600]$.  It is
important to note that for the excitatory synaptic weights, the lower
order expansions are better predictors of the spike probability for
$w>500$.  Since the product $\mu\hat{p}(1- \hat{p}) = 1/2000$, the
loop-expansion diverges for $w>|\mu\hat{p}(1- \hat{p})|^{-1} = 2000$. 

The observed deviation of the loop-expansion result from the simulation is in a smaller range of the synaptic weights than would be predicted from the radius of convergence of the loop-expansion. The reason for this discrepancy is that there is coupling of higher order moments of the spike probability distribution that were truncated in Eq.\ \ref{eq:expanj}. By adopting a linear approximation of the spike-probability function (Eq.\ \ref{eq:probfnct}), we have lost information about when the higher moments of the distribution become large, as in the case $w<-900$. The linear approximation of the spike probability function also truncates information about  the saturation of the spike probability for $w>500$. On the other hand, in the region of synaptic weight values where the loop-expansion is valid, the spike probability, $P_i(t_{n})$, is in the range $[0.3, 0.8]$, which demonstrates a significant effect of the recurrent connectivity.

We also compared the dependency of the spike probability on the noise of the system with numerical simulations for the case where the  recurrent synaptic weights were set at $w = -500$. We found that the variance of the spike probability covers the prediction for much of the range of noise values. The time average of the spike probability function is not strongly influenced by the noise parameter, and the loop-expansion prediction for  $K$-terms of the expansion (Eq.\ \ref{eq:probloop}) matches the simulation when the noise is high ($1/\mu>300$). The deviation of the prediction becomes prominent where the loop expansion diverges, the noise is low. For much of the parameter range, the higher order terms of the expansion do not greatly improve the prediction of spike probability. This lack of improvement greater predictive accuracy in the high noise is the basis of the mean-field approximation  \cite{Treves93} where only the first two terms of the loop expansion are used.

\subsection{\label{sec:stepstim}Stimulation effects}
We can calculate the effects of surrounding neurons on the response of a single neuron to a current injection by using our expression for the membrane potential of the model neuron. This type of calculation would be useful for quantifying the functional strength of synaptic connections to a neuron embedded in a network when the neuron can only be examined with single cell recordings. If the number of synaptic connections can be estimated using morphological studies, and the noise and baseline activity measured by a single cell recording, then the effects of neighboring neurons on the response of the recorded neuron to a step stimulus can be used to fit the recurrent synaptic strength. 

\begin{figure*}
\includegraphics[width=5in]{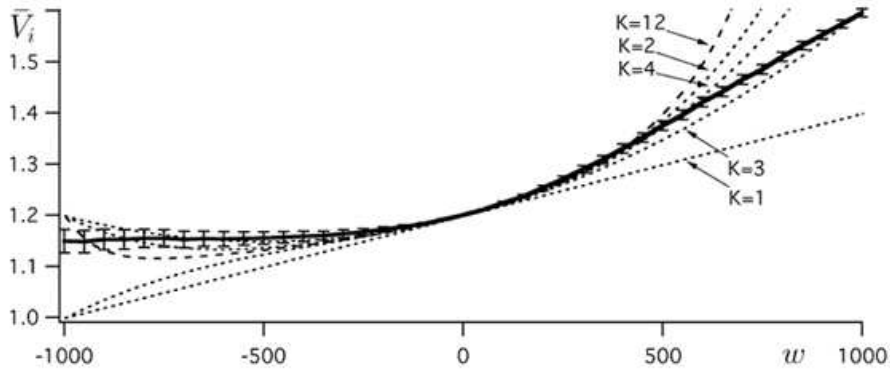}
\caption{\label{fig:potcomps}Comparison of the loop-expansion prediction with simulation for the synaptic weight dependence of the average membrane-potential loop-expansion during a pulsed, intracellular simulation.  The time average of the membrane potential (solid line)  during a repeated step-change of the background input by 20\% of the baseline membrane potential (error bars represent 1 standard deviation). The loop-expansion prediction (dashed lines) for  $K$-terms of the expansion (Eq.\ \ref{eq:probloop}) matches the simulation for weak synaptic weights.  Parameters are as in Fig.\ \ref{fig:probcomps} except the number of time steps for each simulation is $N=2\times 10^5$.}
\end{figure*}

In the simple case of two synaptically coupled neurons, we can represent a constant current injection into one neuron using the following background membrane potentials (from Eq.\ \ref{eq:mempot2})
\be
U_{i} (t_{n}) &=&  \left\{ \begin{array}{ll}   \hat{V} + U_{o} , \mbox{ for  $t_{n}$ during the stimulus,}
\\ \hat{V} \mbox{ otherwise,}  
\end{array}\right.
\nonumber\\
U_{j} (t_{n}) &=& \hat{V} , \mbox{ for all $t_{n}$.} 
\ee
This definition for the background potential alters the our loop expansion of the membrane potential (Eq.\  \ref{eq:mempotexp}). 
Since $U_{i} (t_{n}) - \hat{V} = U_{o}$ , and $U_{j} (t_{n}) - \hat{V} = 0$, we have 
\begin{widetext}
\be
\langle V_{i} (t_{n})\rangle  = \hat{V} + U_{o}  
+ w\hat{p}\sum_{K=1}^{\infty} 
(w\mu \hat{p}(1- \hat{p}))^{K-1}
[1 + \mu(1-\hat{p})\frac{1}{2}(1+(-1)^K)U_{o}],
\ee
\end{widetext}
where we have assumed that the stimulus time is long enough to integrate out the EPSP kernel. 

Comparisons with numerical simulation are shown in Fig.\ \ref{fig:potcomps}. As in previous simulations, we calculated the membrane potential, $V_{i} (t_{n})$,  and calculated spiking of the neuron pair using spike response models. In the simulation, one of two coupled neurons was stimulated with a with a step increase of its baseline membrane potential by 20\% so that $U_{o} = 0.2\hat{V}$. The stimulus was presented every $M$ = 200 time steps and lasted for $L$ = 20 time steps for a total sample time of $N=2\times 10^5$ time steps. The membrane potential was time averaged during the stimulation presentations,
\be
\bar{V_i} = \frac{M}{NL}\sum_{m+1}^{N/M}\sum_{l=1}^{L}V(t_{mM+l}),
\ee 
and the variance was calculated by
\be
\mbox{var}(V_i) = \sqrt{\frac{M}{NL}\sum_{m+1}^{N/M}\sum_{l=1}^{L}(\bar{V_i}-V(t_{mM+l}))^2}.
\ee 
The result show that the loop expansion can predict the response of a neuron embedded in a network with a broad range of synaptic strengths. In these expressions, the results could be generalized to the case where $w_{ij} \neq w_{ji}$ if we recover a sum over the synaptic weights and alter the expressions appropriately.

In the example presented here, we assumed that the stimulation time was long compared to the EPSP so that we integrated out the temporal dependencies for simplicity. However, the change in spike rate caused by the stimulation would provide a transient correlation that persist for the duration of $\epsilon^{(K)}(t_{n})$ \cite{Brunel99,Omurtag00}. This would suggest a mechanism for persistent currents leading to possible memory effects.

\begin{figure*}
\includegraphics[width=5in]{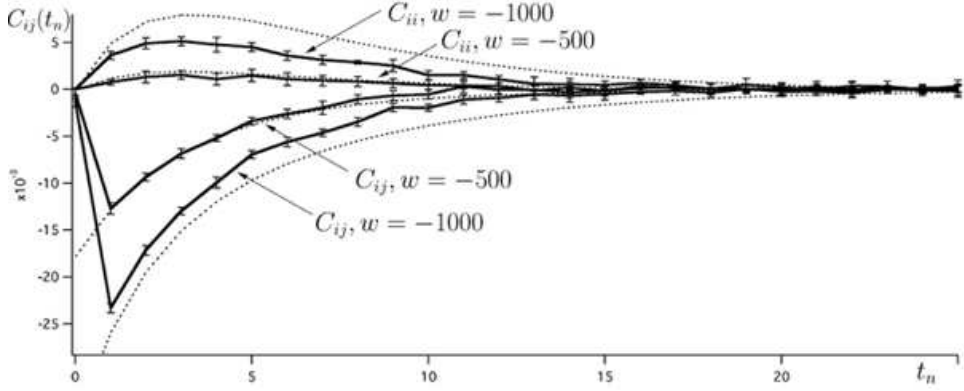}
\caption{\label{fig:correl} Comparison of corelation functions with numerical simulation results. The correlation functions from the simulation were computed by Eq.\  \ref{eq:simcor} and average over 10 trials (solid line, error bars represent 1 standard deviation). The loop-expansion prediction (dashed lines) for a  $6$-term expansion (dashed lines) matches the simulation when the synaptic strength, $w$, was within the valid range found in Fig.\ \ref{fig:probcomps}. The recurrent synaptic contact in the simulation had a 1-step time-delay as is evident in the joint-correlation functions, $C_{ij}$. The  parameters are the same as those used in Fig.\ \ref{fig:probcomps}.}
\end{figure*}

\subsection{\label{sec:correl}Correlation functions}
We can apply our diagrammatic method directly to the problem of calculating correlation functions in a network of coupled neurons. For a pair of neurons, we refer to the loop-expansion diagram (Fig. \ref{fig:loops}), along with the diagrammatic rules, and choose those terms that represent the type of correlation that we wish to calculate. For an auto-correlation function, we use only the even power terms that represent the effect of a neuron on itself through the other neuron in the network, and for a joint-correlation function, we use only the odd power terms the give the influence of one neuron on the other.

\subsubsection{\label{sec:corAnaly}Parameter dependency of correlation 
functions}
We wish to calculate the spike correlation function \cite{Ginzburg94}, 
\be
C_{ij}(s_{n}) &=& \langle (\langle S_i (t_{n})\rangle - S_i (t_{n} ))\times\nonumber \\
&&(\langle S_j (t_{n}+ s_{n}) \rangle - S_j (t_{n}+ s_{n}))\rangle .
\ee
In our example, we have let the background membrane potential be a constant so that $\langle S_i (t_{n}) \rangle $ = $\hat{p}$. To obtain an expression for the auto-correlation function ($j=i$), we multiply the even terms of the loop expansion (Fig. \ref{fig:loops}) by an overall factor of $\hat{p}$ that arises from the product of spike probabilities,
\be
C_{ii}(t_{n}) =  \hat{p}^{2}\sum_{K=1}^{\infty} (w\mu \hat{p}(1- 
\hat{p}))^{2K}\epsilon^{(2K)}(t_{n}).
\label{eq:autocor}
\ee
The terms of the loop-expansion represent the effect of neuron $i$ on itself via propagation $K$ times around the loop. If the background membrane potential were not constant, then there would have been terms in Eq.\ \ref{eq:autocor} that result from the temporal correlations of $U_i(t_n)$.

The joint-correlation function is found by multiplying the odd terms of the loop-expansion by the average spike probability,
\be
C_{ij}(t_{n}) =  \hat{p}^{2}\sum_{K=1}^{\infty} (w\mu \hat{p}(1- 
\hat{p}))^{2K-1}\epsilon^{(2K-1)}(t_{n}).
\label{eq:jcor}
\ee
The first term in the series is the correlation as a result of the direct effect of neuron-$j$ on neuron-$i$, the second term in the series is the effect for the signal that has traveled once around the loop. 

\subsubsection{\label{sec:corNum}Numerical comparison with correlation functions}
In order to test our analytic expressions for the correlation functions of two mutually connected neurons, we simulated a pair of spike-response model neurons and computed the spike correlation functions using the formula:
\be
C_{ij}^{sim}(t_{n}) = \frac{1}{R}\sum_{m=1}^{N} S_i(t_m)S_j(t_m + t_n) - \bar{S_i} \bar{S_j}
\label{eq:simcor}
\ee
where $R$ is the number of spikes in the spike train of neuron-$i$ and $\bar{S_i} $ is the time-averaged spike probability,  $\bar{S_i} = (1/N)\sum_{n=1}^NS_i(t_n)$. The sample time was $N=2\times 10^5$. In the case of auto-correlation functions ($i=j$), we set $C_{ii}^{sim}(0)=0$ \cite{Gerstner02}.
  
A comparison is shown in Fig.\ \ref{fig:correl} for the auto- and joint-correlation functions of neurons coupled by two choices of synaptic weights.  In
Fig.\ \ref{fig:correl} we compare the prediction of the loop-expansion to the average correlation function of 10 simulations. The prediction in the case of the stronger weight is shown to deviate from the simulation due to higher order terms containing higher order convolutions of the EPSP function, $\epsilon^{(K)}(t_{n})$.

\subsection{\label{sec:hightemp}Comparison with the high-temperature expansion}
A similar method of deriving analytic expressions for collective variables in statistical systems was developed that is valid under conditions of high temperatures \cite{Horwitz61,Fisher64,Stanley66}. Such an expansion takes advantage of the noise parameter in our spike probability function (Eq.\ \ref{eq:probfnct}). This function is equivalent to the distribution function of fermionic systems where the parameter $\mu$ is proportional to the inverse temperature of the system. Expanding Eq.\ \ref{eq:probfnct} near $\mu=0$ yields
\be
P_i(t_n) &=& \frac{1}{2}+\frac{1}{4}\mu(V_i(t_n)-\theta_i) - \frac{1}{8}\mu^3(V_i(t_n)-\theta_i)^3
\nonumber \\
&&+\frac{1}{4}\mu^5(V_i(t_n)-\theta_i)^5 +\cdots
\ee
As in our earlier loop-expansion near a fixed point of $P_i(t_n)$, we substitute in the explicit expression for $V_i(t_n)$ which contains a factor representing the spike probability of the neurons connected to neuron-$i$. The spike probability functions for the coupled neurons  are also expanded, and the linear terms are collected so that we arrive at an expression for the spike probability of neuron-$i$. This expansion is the sum of its spike probability in the absence of synaptic connections plus a series of perturbations due to recurrent connections,
\be
P_i(t_n) = P_i^U(t_n) + \sum_{K=1}(\frac{1}{4}\mu w)^K \epsilon^{(K)}*P_a^U(t_n),
\ee
where $a=0$ if $K$ is even, and  $a=1$ if $K$ is odd. The expression obtained from the high-temperature expansion is therefore equivalent to the fixed-point loop-expansion for the special case when the fixed-point $\hat{p}=1/2$. Thus, the fixed-point loop-expansion can be thought of as a high-temperature expansion with a bias and with a temporal structure in the interaction kernels. 

Although the fixed-point loop-expansion was developed for biological systems that possess homeostatic adaptation mechanisms, there are advantages to this connection with the high-temperature expansion. First, for more complicated networks, the diagrams that weight the terms of the perturbation expansion have already been enumerated in the literature \cite{Stanley67}. Second, the connection may be made with other techniques from statistical mechanics \cite{Shankar94} to calculate properties of neural networks that could be used to test models of biological neural systems.

\section{\label{sec:multiple}Networks with multiple neurons}
\subsection{\label{sec:chains}Effective spike propagation\\
Diminishing efficacy and equivalent networks}

The propagation of spikes across multiple neurons in a neural network can be estimated by the loop-expansion of the correlation function for a periodic chain of neurons \cite{Abeles91}, where each neuron is recurrently coupled to its nearest neighbors and the first neuron is coupled to the last. When more than two neurons are on a network, the number of loop diagrams that contribute to each term of the loop-expansion must be enumerated. 

\begin{figure}
\includegraphics{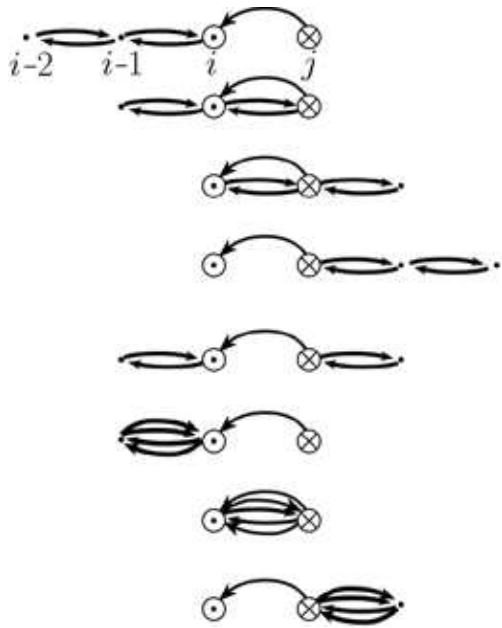}% Here is how to import EPS art
\caption{\label{fig: chainLoops}Loop diagrams for the $K=5$ term in the correlation function, $C_{ij}(t_{n})$, of a periodic chain of neurons. There are 5 synaptic links with two loops. The computation is symmetric under the transposition of $i$ and $j$. }
 \end{figure}

Consider the joint-correlation function between a neighboring pair of neurons in the network depicted in Fig.\ \ref{fig:model}. The terms of the correlation function (Eq.\ \ref{eq:jcor}) must by weighted by the number of distinct loop diagrams, $A_K^l$ that can be constructed with $(K-l)/2$ loops, where $l$ is the number of synaptic links that separate the neurons ($j=i\pm l$). Then we have
\be
C_{ij}(t_{n}) =  \hat{p}^{2}\sum_{K=1}^{\infty} A_K^l (w\mu \hat{p}(1- 
\hat{p}))^{2K-1}\epsilon^{(2K-1)}(t_{n}).
\ee
For nearest neighbors ($j=i+1$), the coefficients are: $A_1^1 = 1$, $A_3^1 = 3$, $A_5^1 = 8$. The distinct loops contributing to $A_5$ are shown in Fig.\ \ref{fig: chainLoops}.

\begin{figure*}
\includegraphics[width=6in]{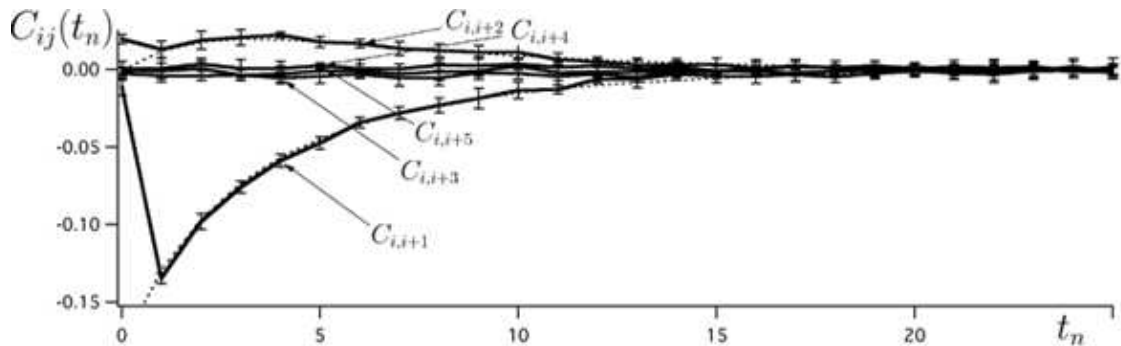}
\caption{\label{fig:correlSpace} Diminishing efficacy and equivalent networks.  The joint-correlation function of neuron pairs that are separated by more than 2 synaptic connections are indistinguishable from noise. A comparison of joint-correlation functions with numerical simulation results for $C_{i,i+1}$ and $C_{i,i+2}$ demonstrate that the loop-expansion generalizes to networks of multiple neurons. The  parameters are the same as those used in Fig.\ \ref{fig:probcomps} with the synaptic weights set at $w = -500$.}
\end{figure*}

We compared the prediction of the loop-expansion for the joint-correlation function with a simulated periodic chain network of 10 neurons (Fig.\ \ref{fig:correlSpace}).  The model parameters were the same as used to generate Fig.\ \ref{fig:probcomps}, with the synaptic weights set at $w = -500$. We averaged over 10 simulations and the number of time steps in each simulation was $N=2\times 10^5$. The results are shown in Fig.\ \ref{fig: chainLoops} where the joint-correlation function was computed in the simulation for 5 different locations. The loop-expansion was computed for the nearest neighbor ($j=i+1$), and the neuron two steps away ($j=i+2$). The joint-correlation function of the neuron pair that are separated by 2 synaptic links ($j=i+2$), has coefficients: $A_2^2 = 1$, $A_4^2 = 4$, $A_6^2 = 12$. The joint-correlation functions for neuron pairs separated by more than 2 synaptic links  were indistinguishable from the noise.

The result that neurons located further than two steps away had a vanishing correlation function suggests a rapid, diminishing efficacy of one neuron's synaptic action on other neurons in the network. This diminishing efficacy is in a parameter range where network recurrence reduces the spike probability of each neuron by 20\% (see Fig.\ \ref{fig:probcomps}). The signals that may begin with one of the neurons do not propagate far, even though the local connectivity has a strong effect on each individual neuron.

Since the correlation function is indistinguishable from zero after 3 synaptic links, we could have derived identical results with a periodic chain network of only 5 neurons. Thus large networks may be simulated by smaller, equivalent networks to yield the same results for neuron activity variables such as spike probability and correlation functions.

\subsection{\label{sec:contin} Continuum limit of spatial components }
In large biological neural networks, identifying and analyzing the multiple connections between a large number of neurons neurons can be cumbersome. A mean-field approach has been developed \cite{Wilson73,Gerstner95} to deal with large populations of interacting neurons where the neural populations are represented by a continuous field. In our notation, the spike-probability function will be generalized to include spatial components, $P({\bf x}, t_n)$, where ${\bf x} \in {\mathbb R}^n$ and $1\leq n \leq3$. The spike probability at each point in the network has the same dependence on the generalized membrane potential, $V({\bf x}, t_n)$, as in the discrete neuron case (Eq.\ \ref{eq:probfnct}), but now the neurons are labeled by their spatial location, ${\bf x}$. 

Synaptic interactions are introduced by extending the synaptic weights to a synaptic density, $w({\bf x})$, so Eq.\ \ref{eq:mempot} becomes
\be
\langle V({\bf x}, t_{n})\rangle &=& U ({\bf x}, t_{n}) \\
&&+ \int d{\bf x}' \sum_{m} w({\bf x}-{\bf x}')\epsilon(t_{m}-t_{n})P({\bf x}', t_{m})\nonumber
\label{eq:mempotCont}
\ee
We can combine the synaptic density with the PSP kernel to define an effective PSP kernel that depends on both space and time, $W({\bf x}, t_{n}) = w( {\bf x})\epsilon(t_{n})$.
Repeating our calculations from the discrete case (Section \ref{sec:pair}) we arrive at the loop expansion for the spike probability of a continuous neural network
\be
P({\bf x}, t_{n}) = \sum_{K=0}^{\infty} (\mu \hat{p}(1- \hat{p}))^{K}
W^{(K)}*P^U({\bf x}, t_{n})
\label{eq:probloopCont}
\ee
where the convolution is now defined over spatial as well as temporal variables, 
\begin{widetext}
\be
W^{(K)}*f({\bf x}, t_n) =
\int d{\bf x}_{1}\sum_{i_{1}}\cdots\int d{\bf x}_{K}\sum_{i_{K}}w( {\bf x}_{K-1}- {\bf x}_{K})\epsilon(t_{i_{K-1}}-t_{i_{K}})\cdots
w( {\bf x}- {\bf x}_{1})\epsilon(t_{n}-t_{i_{1}})f({\bf x}, t_{i_1}),
\ee
\end{widetext}
where all $n$ components of ${\bf x}$ are integrated.

The loop corrections to any mean-field model of a large neural network can now be computed for the case where each location in the network is driven by a different input as represented by the background spike probability, $P^U({\bf x}, t_{n})$. Although we have only included one type of neuron in the above calculations, we may further generalize the PSP kernel to couple different types of neurons, as we will demonstrate in the next section.

\section{\label{sec:appl}Applications to biological networks}
An important application of the techniques presented here is to biological neural networks found in laminar structures with lateral synaptic connectivity such as the mammalian cerebral cortex and the cerebellum. In the cerebral cortex, pyramidal cells are coupled with lateral excitatory synapses. In addition, there are inhibitory interneurons that are excited by, and inhibit, pyramidal cells. Interacting pools of excitatory and inhibitory neurons have been modeled using mean-field approaches \cite{Wilson72} and have been used to study instabilities under pathological conditions in the visual cortex  \cite{Ermentrout79}. We will extend these previous results because we include the time-dependence in the synaptic kernel function and we add perturbative corrections due to recurrence. 

The synaptic densities of the two neural populations are modeled with a Gaussian function  \cite{Ermentrout79},
\be
w_a ({\bf x}) = \frac{w_a}{\sigma_a\sqrt{\pi}} e^{-({\bf x}/\sigma_a)^2},
\ee
where $a$ denotes whether the synapses are excitatory ($a=E$) or inhibitory ($a=I$), $w_a$ scales the synaptic strength, and $\sigma_a$ is the lateral extent of the synaptic coupling. In the 2-dimensional cerebral cortex, ${\bf x} \in {\mathbb R}^2$ and ${\bf x}^2 = {\bf x}\cdot{\bf x}$. 

The effect of excitatory pyramidal cells on other pyramidal cells is both excitatory and via interneurons, so we may eliminate explicit reference to the inhibitory population by incorporating the inhibitory interneurons into the definition of the excitatory synaptic density,
\be
W ({\bf x}, t_n) = W_E ({\bf x}, t_n) - W_I ({\bf x}, t_n).
\ee
where the first term represents the excitatory component of the synaptic density, and the second term represents the inhibitory component. 
The time dependency of the inhibitory PSP kernel can be approximated by convolving two PSP kernels because the pyramidal cells inhibit other pyramidal cells through 2 synaptic links. Thus, the form of the synaptic  density becomes
\be
W ({\bf x}, t_n) = w_E ({\bf x}) \epsilon (t_n)  - w_I ({\bf x}) \epsilon^{(2)} (t_n).
\ee
Using this expression for the synaptic density, we may compute the loop expansion for the spike-probability function (Eq.\ \ref{eq:probloopCont}),
\begin{widetext}
\be
P({\bf x}, t_{n}) = \sum_{K=0}^{\infty} (\mu \hat{p}(1- \hat{p}))^{K}
\sum_{l=0}^{K} (-1)^{K-l} \left(\begin{array}{cc} K \\ l \end{array}      \right)
(w_E^{(l)}*w_I^{(K-l)}*)\epsilon^{(2K-l)}*P^U({\bf x}, t_{n}).
\label{eq:probloopCtx}
\ee
\end{widetext}
Since each convolution of Gaussian functions increases the standard deviation, each higher order term of the loop extends the reach of the synaptic densities. Thus, as $K$ increases in the loop expansion, each term predicts the distance across the cortex that neurons contribute to the spike probability of a given neuron. This expression for the spike-probability function could be improved by including more details about the inhibitory interneurons, such as by taking care to include an independent fixed point spike probability, $\hat{p}$, for each neuronal type.

The validity of the loop expansion depends on whether  the dynamical solution is stable in both the time and space domain for physiological parameter settings of the synaptic weights and noise parameter. Previous studies of cortical dynamics have suggested that spatial instabilities are the result of a pathological state of the system   \cite{Ermentrout79}. We may therefore assume that, under normal physiological conditions, the loop expansion provides a valid, analytic expression for the neural dynamics of the cerebral cortex. 

A possible use of this calculation would be to calculate the activity of pyramidal cells in the primary visual cortex that are driven by sensory information from the lateral geniculate nucleus \cite{Kang03}, represented by $P^U({\bf x}, t_{n})$. The prediction of Eq.\ \ref{eq:probloopCtx}  combines the visual input with the lateral connections within the primary visual cortex. Discrepancies between the predicted spike activity and spike activity measured in experiments would suggest the influence of feedback recurrence from other regions of the visual cortex \cite{Tsodyks95}. Therefore, the loop-expansion technique could be a theoretical tool to study the consequances of synaptic communication between cortical regions.

A computation could also be made to predict the spike correlation between neurons as a function of their separation. Since this analytic technique provides the parameter dependencies of correlation functions, it is possible to deduce the effective synaptic coupling between neurons using data that measures the lateral spread of  pyramidal cell axons and pair recordings to measure the space dependent correlation functions. These calculations would reveal what dynamical state of the cortex (coherent or asynchronous) was present under physiological conditions, but such calculations are beyond the scope of the present  study. 

Another application to biological networks is to predict neural dynamics in the presence of recurrent feedback loops, as found in the mammalian thalamic-cortical system and the cerebellum. In the cerebellum, the mossy fiber input to the system is controlled by recurrent inhibition from inhibitory interneurons that appear to control the processing of information carried by mossy fibers. If the recurrent inhibition plays a modulatory role that can be modeled as a perturbation of the system, then this technique could be valuable to investigate the effects on sensory processing in the granule cell layer of the cerebellum \cite{Roberts03}. 

A class of biological neural networks that falls outside the validity of the loop-expansion as presented here is the  pattern generators such as those found in invertebrate motor systems and the spinal chord. Since these neural systems operate in a bursting, coherent state, the perturbation expansion would be a poor predictor of the neural dynamics. Techniques from dynamical systems would be more appropriate for the study of these systems.

Since the loop-expansion technique is valid when the recurrence plays a modulatory role in a weak and noisy system, the method may be generalized to analyze modulatory biochemical networks. The localized chemical concentrations could be represented by neurons and the kinetics of the biochemistry represented by PSP kernels.  The results could predict dynamics of complex biochemical networks where many components interact. 

In summary, we have introduced an analytic method to compute dynamical variables that predict the spiking activity of neurons embedded in recurrent neural networks. The method relies on an expansion of recurrent synaptic connections and is valid for neural networks in the asynchronous state where there are no global instabilities. A diagrammatic method has been introduced to help construct the terms of the loop-expansion, and the results have been compared with simulations of spiking neurons. The loop-expansion technique is designed to be applied to biological neural networks of spiking neurons, and includes the temporal aspects of synaptic transmission and spike generation. Finally, we demonstrated how the loop-expansion technique can be applied to neural dynamics in the cerebral cortex to extend previous modeling studies.

\begin{acknowledgments}
I would like to thank Alan Williams for helpful comments and edits.
This material is based upon work supported by the National Institutes of Health under Grant No. R01-MH60364 and by the National Science Foundation under Grant No. PHY99-07949. I wish to acknowledge the hospitality of the Institute for Theoretical Physics, Santa Barbara, where much of this work was developed during the program on Dynamics of Neural Networks.
\end{acknowledgments}

\appendix
\section{Spike Probability Functions}
%\subsection{\label{app:subsec}A subsection in an appendix}
In order to develop a mechanistic understanding of noise in our threshold model of a neuron, we will compare three models on neuronal variability. The objective of our model presented in Section \ref{sec:SRN} is to separate the correlated synaptic inputs from uncorrelated noise. The uncorrelated noise consists of random excitatory and inhibitory synaptic inputs, as well as channel noise. These inputs are uncorrelated with the inputs described by the function $V_i(t_n)$ in Eq.\ (\ref{eq:probfnct}). The model neuron described by  Eq.\ (\ref{eq:probfnct}) is an approximation of the Gaussian limit of the Stein model \cite{Gerstein64,Stein65} that is developed from a stochastic differential equation for the membrane potential $v(t)$,
\be
\frac{d}{dt}v(t) = -\frac{v(t)}{\tau} + \xi(t),
\ee
where $\tau$ is the membrane time constant and $\xi(t)$ represents uncorrelated Gaussian noise with zero mean: $\langle \xi(t)\rangle = 0$ and $\langle \xi(t)\xi(t')\rangle = \sigma^2\delta(t-t')$. In Stein's model, $\xi(t)$ represents inputs from balanced excitatory and inhibitory synaptic noise.

\begin{figure*}
\includegraphics[width=6in]{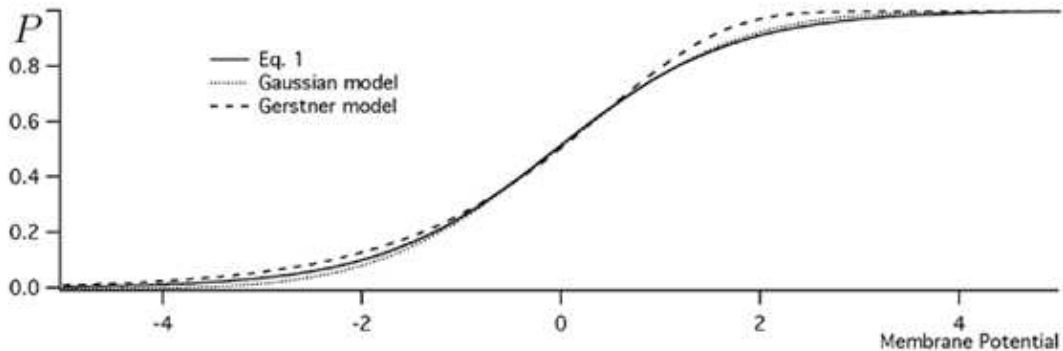}
\caption{\label{fig_models} Comparison of three forms of spike probability functions. The solid trace is the function defined in Eq.\ (\ref{eq:probfnct}), the dotted trace is the function defined in Eq.\ (\ref{eq:gauss}), the dashed trace is the function defined in Eq.\ (\ref{eq:gerstner}). The noise parameters $\sigma = 0.5$, $\mu = 4\sigma/\sqrt{\pi}$, and $\beta = \mu/\ln(4)$, and the threshold, $\theta=0$. The differences between these models are within the variability of the simulations compared with the loop expansion in previous figures.}
\end{figure*}

In the diffusion limit of many small synaptic inputs, it can be shown \cite{Lansky84} that Stein's  model approximates an Ornstein-Uhlenbeck process \cite{Uhlenbeck30} with a stationary membrane potential that takes the form of a Gaussian function,
\be
\Phi(v, \sigma) = \frac{1}{\sigma\sqrt{\pi}}\exp\left[-\left(\frac{v-V}{\sigma}\right)^2\right],
\ee
where $V$ is the stationary mean of $v(t)$. A neuron with a membrane potential described by this Gaussian distribution, and a spike threshold of $\theta$, has the spike probability function,
\be
P_{Gauss} (V, \theta) = \int_{\theta}^{\infty} \Phi(v, \sigma)\, dv.
\label{eq:gauss}
\ee
To compare this spike probability function with Eq. (\ref{eq:probfnct}), we equate the slope of $P_{Gauss} (V, \theta)$ with the slope of $P_i (V_i)$ at $V_i=\theta$ to arrive at $\mu = 4\sigma/\sqrt{\pi}$. A comparison of the functional form of Eq.\ (\ref{eq:gauss}) and Eq.\ (\ref{eq:probfnct})  is shown in Fig.\ \ref{fig_models}. In our treatment, we embed the effects of uncorrelated synaptic noise as background noise into the spike probability function (Eq.\ \ref{eq:probfnct}). The source of noise parametrized by $\mu$ is uncorrelated with the synaptic input from other neurons explicitly represented in the network and with the background input, $\langle \xi(t)U_i(t')\rangle = 0$. Improvements to this approximation could include more details of how synaptic dynamics affect the characteristics of the noise \cite{Fourcaud02}.

It is instructive to compare the spike probability function developed by Gerstner \cite{Gerstner93, Gerstner95}, because Gerstner's model has an explicit expression for the instantaneous spike-rate function, $\rho(V)$, based on an Arrhenius hazard function \cite{Plesser00}. Our spike probability function given by  Eq.\ (\ref{eq:probfnct})  estimates the probability of a spike in the interval $\tri t$. The sigmoid threshold function is similar to Gerstner's interval spike probability function found by integrating the probability of surviving without a spike, $\exp(-t \rho(V))$, over a time interval. The instantaneous spike-rate is $\rho(V) = r_0\exp[\beta(V-\theta)]$, where $\beta$ is the noise parameter and $r_0$ is the  spike rate at threshold. Integrating $\exp(-t \rho(V))$ over the interval $\tri t$ yields the spike probability function \cite{Gerstner92}
\be
P_{Gerstner} (V, \theta) = 1-\exp\left[- r_0\tri t\exp[\beta(V-\theta)]\right].
\label{eq:gerstner}
\ee
Equating this function with our spike probability functions at $V=\theta$ (assuming $\tri t = 1$ ms) allows us to compute $r_0=\ln(2)$. If we expand $P_{Gerstner} (V, \theta)$ about  $V=\theta$, we find that the relationship between  the noise functions is $\mu = \ln(4)\beta$. Gerstner's spike probability function is compared with the other two probability functions in Fig.\ \ref{fig_models}.

%\bibliography{references}% Produces the bibliography via BibTeX.

\end{document}